\documentclass[11pt,a4paper]{article}
\usepackage{enumerate, amsmath, amsthm, mathrsfs, array, amssymb, indentfirst, latexsym, pstricks}
\allowdisplaybreaks[1]

\newtheoremstyle{bthm}{\baselineskip}{\baselineskip}{\slshape}{}{\bfseries}{:}{\newline }{}
\newtheoremstyle{bex}{\baselineskip}{\baselineskip}{}{}{\sffamily}{:}{\newline }{}
\theoremstyle{bthm}
\theoremstyle{bex}
\newtheorem{myth}{Theorem}
\newtheorem{mylm}{Lemma}
\newtheorem{cor}{Corollary}
\begin{document}
\title{Wiener index of binomial trees and Fibonacci trees}
\author{K. Viswanathan Iyer \thanks{Corresponding author} \\ Department of Computer Science and Engineering\\
National Institute of Technology\\ Tiruchirapalli--620 015, India\\
email: kvi@nitt.edu 
\and
Udaya Kumar Reddy K.R \\ Department of Computer Science and Engineering\\ National Institute of Technology\\ Tiruchirapalli--620 015, India\\ email: krudaykumar@yahoo.com 
}
\date{}
\maketitle
\thispagestyle{empty}
\begin{abstract}
Given a simple connected undirected graph $G$ = ($V$, $E$), the $Wiener$ $index$ $W$($G$) of $G$ is defined as half the sum of the distances $d$($u$, $v$) between all pairs of vertices $u$, $v$ of $G$, where $d$($u$, $v$) denotes the distance (the number of edges on a shortest path between $u$ and $v$) between $u$, $v$ in $G$. We obtain an expression for $W$($G$), where $G$ is a binomial tree. For Fibonacci trees and binary Fibonacci trees with $F_k$ ($k$-th Fibonacci number) vertices, we outline algorithms for computing their Wiener index in time $O$($\log{F_k}$). This may be compared with the existing result: for any tree $T$ with $n$ vertices, $W$($T$) can be computed by an algorithm in time $O$($n$).
\begin{quote}
$Keywords$: binomial tree, binary Fibonacci trees, distance in graphs, Fibonacci trees, Wiener index. 
\end{quote}
\end{abstract}

\section{Introduction}
Let $G$ = ($V$($G$), $E$($G$)) be a connected undirected graph without self-loops and multiple edges. The $Wiener$ $index$ $W$($G$) of $G$ is defined as
\begin{equation}
W(G) =
\frac{1}{2} 
\sum_{u \in V(G)} 
\sum_{v \in V(G)}
d(u, v), \label{name1}
\end{equation}
where $d$($u$, $v$) denotes the shortest distance between $u$, $v$ in $G$.\\
\indent Wiener index is a distance based graph invariant which is one of the most popular topological indices in mathematical chemistry. It is named after the chemist Harold Wiener, who first introduced it in 1947 to study chemical properties of alkanes. It is now recognized that there are good correlations between $W$($G$) and physico-chemical properties of the organic compound from which $G$ is derived, especially when $G$ is a tree. As mentioned in \cite{wag1}, in the process of drug design, one may want to construct chemical compounds from common molecules so that the resulting compound has an expected Wiener index. For chemical applications of Wiener index, see \cite{klav, wag1}. For more details on Wiener index of trees and their applications, see \cite{dob}. \\
\indent We begin with the definitions of binomial trees and two types of Fibonacci trees. Let $k$ be a nonnegative integer. The $binomial$ $tree$ $T_{B_k}$ of order $k$, is an ordered tree defined recursively in the following way: 
\begin{itemize}
\item[i.] The binomial tree $T_{B_0}$ consists of a single node.
\item[ii.] For $k$ $\geq$ 1, the binomial tree $T_{B_k}$ is constructed from two binomial trees $T_{B_{k-1}}$ of order $k-1$ by attaching one of them as the leftmost child of the root of the other.
\end{itemize}
For more details on binomial trees and its properties, see \cite{CLRS}. Fig.~\ref{fig1}(a) shows the binomial trees $T_{B_0}$ through $T_{B_3}$ and Fig.~\ref{fig1}(b) shows the binomial tree $T_{B_k}$. Let $F_k$ denote the $k$-th Fibonacci number. The $Fibonacci$ $tree$ $T_{f_k}$ of order $k$, is defined recursively in the following way: 
\begin{itemize}
\item[i.] $T_{f_{-1}}$ and $T_{f_0}$ are both Fibonacci trees consisting of a single node. 
\item[ii.] For $k$ $\geq$ 1, $T_{f_k}$ consists of two Fibonacci trees $T_{f_{k-1}}$ and $T_{f_{k-2}}$ of orders $k-1$, and $k-2$, respectively, where $T_{f_{k-2}}$ is the rightmost child of the root of the the other.
\end{itemize}
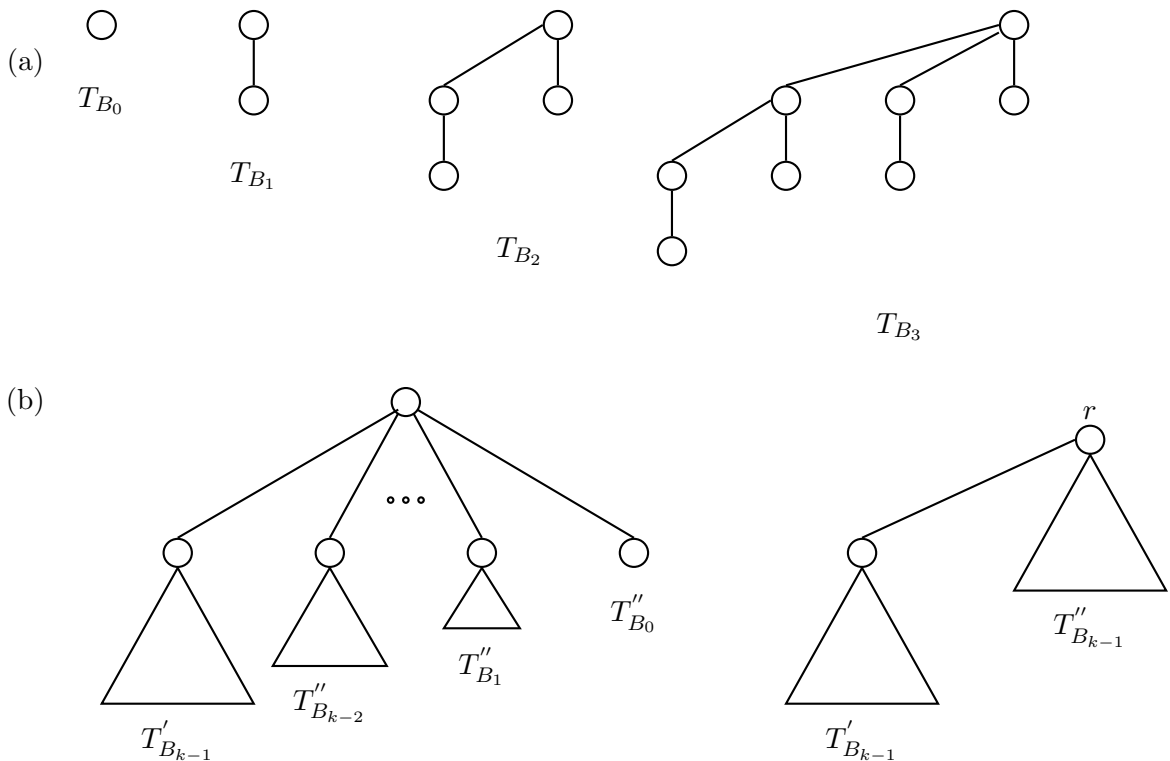
\begin{figure}[tb]
\centering
\begin{pspicture}(1,2.5)(16,12)
\pscircle(2,12){0.2}
\pscircle(4,12){0.2}
\pscircle(4,11){0.2}
\psline(4,11.2)(4,11.8)
\pscircle(8,12){0.2}
\pscircle(8,11){0.2}
\psline(8,11.2)(8,11.8)
\pscircle(6.5,11){0.2}
\pscircle(6.5,10){0.2}
\psline(7.8,12)(6.5,11.2)
\psline(6.5,10.2)(6.5,10.8)
\pscircle(14,12){0.2}
\pscircle(14,11){0.2}
\pscircle(12.5,11){0.2}
\pscircle(12.5,10){0.2}
\pscircle(11,11){0.2}
\pscircle(11,10){0.2}
\pscircle(9.5,10){0.2}
\pscircle(9.5,9){0.2}
\psline(14,11.2)(14,11.8)
\psline(12.5,10.2)(12.5,10.8)
\psline(11,10.2)(11,10.8)
\psline(9.5,9.2)(9.5,9.8)
\psline(12.5,11.2)(13.8,11.9)
\psline(11,11.2)(13.8,12)
\psline(9.5,10.2)(10.8,11)
\rput(1,11.5){(a)}
\rput(15,6.85){$r$}
\rput(2,11){$T_{B_0}$}
\rput(4,10){$T_{B_1}$}
\rput(7.5,9){$T_{B_2}$}
\rput(12.5,8){$T_{B_3}$}
\rput(1,7){(b)}

\pscircle(3,5){0.2}
\pscircle(5,5){0.2}
\pscircle(7,5){0.2}
\pscircle(9,5){0.2}
\pscircle(6,7){0.2}
\pspolygon(2,3)(4,3)(3,4.8)
\pspolygon(4.25,3.5)(5.75,3.5)(5,4.8)
\pspolygon(6.5,4)(7.5,4)(7,4.8)
\pscircle(5.8,5.7){0.05}
\pscircle(6,5.7){0.05}
\pscircle(6.2,5.7){0.05}
\psline(3,5.2)(5.9,6.9)
\psline(5,5.2)(5.9,6.85)
\psline(7,5.2)(6.1,6.85)
\psline(9,5.2)(6.15,6.9)
\rput(3,2.5){$T^{'}_{B_{k-1}}$}
\rput(5,3){$T^{''}_{B_{k-2}}$}
\rput(7,3.5){$T^{''}_{B_1}$}
\rput(9,4.2){$T^{''}_{B_0}$}

\pscircle(12,5){0.2}
\pscircle(15,6.5){0.2}
\pspolygon(11,3)(13,3)(12,4.8)
\pspolygon(14,4.5)(16,4.5)(15,6.3)
\psline(12,5.2)(14.8,6.5)
\rput(12,2.5){$T^{'}_{B_{k-1}}$}
\rput(15,4){$T^{''}_{B_{k-1}}$}
\end{pspicture}
\caption{Binomial trees. (a) The binomial trees $T_{B_0}$ through $T_{B_3}$. (b) Two ways of looking at the binomial tree $T_{B_k}$ (Note: $T_{B_{k-1}}$ $\simeq$ $T^{'}_{B_{k-1}}$ $\simeq$ $T^{''}_{B_{k-1}}$). }
\label{fig1}
\end{figure}
Denote $\mid$$T$$\mid$ = number of vertices in the tree $T$. We have $\mid$$T_{f_k}$$\mid$ = $\mid$$T_{f_{k-1}}$$\mid$ + $\mid$$T_{f_{k-2}}$$\mid$, where $\mid$$T_{f_k}$$\mid$ = $F_{k+2}$. For more details on Fibonacci trees, see \cite{BB}. Fig.~\ref{fig2}(a) shows the Fibonacci trees $T_{f_{-1}}$ through $T_{f_4}$ and Fig.~\ref{fig2}(b) shows the Fibonacci tree $T_{f_k}$. Let $T^{b}_{f_k}$ denote the binary Fibonacci tree of order $k$. $T^{b}_{f_k}$, is a rooted tree defined recursively in the following way: 
\begin{itemize}
\item[i.] $T^{b}_{f_0}$ and $T^{b}_{f_1}$ are both the rooted tree consisting of no nodes and a single node respectively. 
\item[ii.] For $k$ $\geq$ 2, the rooted tree $T^{b}_{f_k}$ is constructed from a root with two Fibonacci trees $T^{b}_{f_{k-1}}$ (of order $k-1$) as its left subtree and $T^{b}_{f_{k-2}}$ (of order $k-2$) as its right subtree. 
\end{itemize}
It follows $\mid$$T^{b}_{f_k}$$\mid$ = $\mid$$T^{b}_{f_{k-1}}$$\mid$ + $\mid$$T^{b}_{f_{k-2}}$$\mid$ + 1, where $\mid$$T^{b}_{f_k}$$\mid$ = $F_{k+2}$ $-$ 1. For any vertex $v$ $\in$ $T^{b}_{f_k}$, $deg(v) \leq 4$---hence the binary Fibonacci trees are chemical trees. For more details on binary Fibonacci trees, see \cite{wagner}. Fig.~\ref{fig3}(a) shows the binary Fibonacci trees $T^{b}_{f_0}$ through $T^{b}_{f_4}$ and Fig.~\ref{fig3}(b) shows the binary Fibonacci tree $T^{b}_{f_k}$.\\ We need the following lemma and its corollary (see \cite{dob}).
\begin{figure}[tb]
\centering
\begin{pspicture}(1,5)(16,12)
\pscircle(2,12){0.2}
\pscircle(3.5,12){0.2}
\pscircle(5,12){0.2}
\pscircle(5,11){0.2}
\psline(5,11.2)(5,11.8)
\pscircle(6.75,12){0.2}
\pscircle(7.5,11){0.2}
\psline(6.1,11.15)(6.6,11.8)
\pscircle(6,11){0.2}
\psline(7.4,11.15)(6.8,11.9)
\pscircle(8.5,11){0.2}
\pscircle(9.75,11){0.2}
\pscircle(9.75,12){0.2}
\psline(8.6,11.2)(9.6,11.9)
\psline(9.7,11.2)(9.7,11.8)
\pscircle(11,11){0.2}
\pscircle(11,10){0.2}
\psline(11,10.2)(11,10.8)
\psline(11,11.2)(9.9,11.9)
\pscircle(12,11){0.2}
\pscircle(13,11){0.2}
\pscircle(14,11){0.2}
\pscircle(15.25,11){0.2}
\pscircle(13.5,12){0.2}
\pscircle(14,10){0.2}
\pscircle(14.75,10){0.2}
\pscircle(15.75,10){0.2}
\psline(12,11.2)(13.3,11.9)
\psline(13,11.2)(13.4,11.8)
\psline(14,11.2)(13.6,11.8)
\psline(15.2,11.2)(13.6,11.9)
\psline(14,10.2)(14,10.8)
\psline(14.8,10.2)(15.1,10.9)
\psline(15.4,10.9)(15.7,10.2)
\pscircle(5,8){0.2}
\pscircle(8.25,7){0.2}
\psline(5.2,8)(8.2,7.2)
\pspolygon(5,7.8)(4,6)(6,6)
\pspolygon(7.75,5.5)(8.75,5.5)(8.25,6.8)
\rput(1,11.5){(a)}
\rput(5,8.4){$r$}
\rput(2,11){$T_{f_{-1}}$}
\rput(3.5,11){$T_{f_0}$}
\rput(5,10){$T_{f_1}$}
\rput(6.8,10){$T_{f_2}$}
\rput(9.75,9){$T_{f_3}$}
\rput(14,9){$T_{f_4}$}
\rput(5,5.5){$T_{f_{k-1}}$}
\rput(8.3,4.8){$T_{f_{k-2}}$}
\rput(1,8){(b)}
\end{pspicture}
\caption{Fibonacci trees. (a) Fibonacci trees $T_{f_{-1}}$ through $T_{f_4}$. (b) Fibonacci tree $T_{f_k}$.}
\label{fig2}
\end{figure}
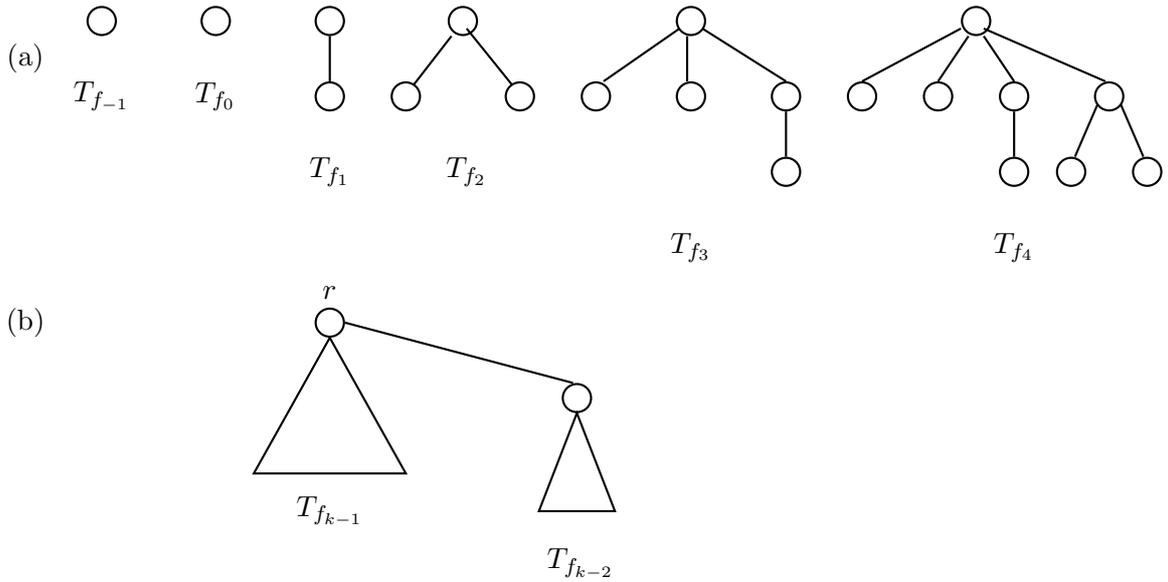

\begin{figure}[tb]
\centering
\begin{pspicture}(1,7)(17,12)
\pscircle(2.5,12){0.2}
\pscircle(3.5,11){0.2}
\pscircle(4.5,12){0.2}
\psline(3.6,11.1)(4.4,11.9)
\pscircle(6,11){0.2}
\pscircle(7,12){0.2}
\psline(6.1,11.1)(6.9,11.9)
\pscircle(8,11){0.2}
\pscircle(5,10){0.2}
\psline(7.1,11.9)(7.9,11.1)
\psline(5.9,10.9)(5.1,10.1)

\pscircle(10,11){0.2}
\pscircle(11,12){0.2}
\psline(10.1,11.1)(10.9,11.9)
\pscircle(12,11){0.2}
\pscircle(9,10){0.2}
\psline(11.1,11.9)(11.9,11.1)
\psline(9.1,10.2)(9.9,10.9)

\pscircle(10.65,10){0.2}
\pscircle(11.35,10){0.2}
\pscircle(8,9){0.2}
\psline(10.6,10.2)(10.1,10.8)
\psline(8.15,9.2)(8.9,9.9)
\psline(11.4,10.2)(11.9,10.9)
\rput(9,7){(a)}
\rput(2.5,11){$T^{b}_{f_1}$}
\rput(4,10){$T^{b}_{f_2}$}
\rput(6.5,9){$T^{b}_{f_3}$}
\rput(10.75,8){$T^{b}_{f_4}$}
\rput(15,7){(b)}

\pscircle(14.5,12){0.2}
\pscircle(13.5,11){0.2}
\pscircle(15.5,11){0.2}
\psline(13.5,11.2)(14.4,11.9)
\psline(15.5,11.2)(14.65,11.9)
\pspolygon(12.5,9)(14.5,9)(13.5,10.8)
\pspolygon(15,9.5)(16,9.5)(15.5,10.8)
\rput(15,12.1){$r$}
\rput(13.3,11.4){$r^{'}$}
\rput(13.5,8.4){$T^{b}_{f_{k-1}}$}
\rput(15.5,8.8){$T^{b}_{f_{k-2}}$}
\end{pspicture}
\caption{Fibonacci trees represented in the form of a binary tree. (a) The binary Fibonacci trees $T^{b}_{f_0}$ through $T^{b}_{f_4}$. (b) The binary Fibonacci tree $T^{b}_{f_k}$.}
\label{fig3}
\end{figure}
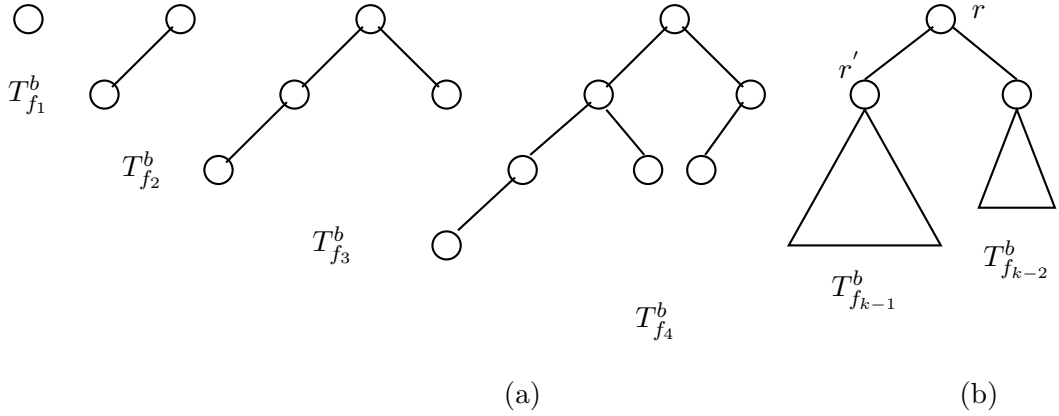
\begin{mylm}
\label{lem1}
\it Let $T_1$ and $T_2$ be two trees with $\mid$V$($$T_1$$)$$\mid$ = $n_1$ and $\mid$V$($$T_2$$)$$\mid$ = $n_2$ respectively, such that V$($$T_1$$)$ $\cap$ V$($$T_2$$)$ = $\phi$. Let T be a tree obtained by identification of a vertex of $T_1$ and a vertex of $T_2$. Let this common vertex be u.  Then
\begin{align}
W(T) = W(T_1) + W(T_2) + (n_1 - 1)D_{T_2}(u) + (n_2 - 1)D_{T_1}(u),
\end{align}
where, for any vertex v in T, $D_T$$($v$)$ denotes the sum of the distances of the form d$($v, x$)$, for all $x$ in V$($T$)$. 
\end{mylm}
\em Proof. \rm Follows from a simple combinatorial argument. \qed \\
The following corollary is a consequence of Lemma~\ref{lem1}.
\begin{cor}
\label{cor1}
\it Let $T_1$ and $T_2$ be two trees with $\mid$V$($$T_1$$)$$\mid$ = $n_1$ and $\mid$V$($$T_2$$)$$\mid$ = $n_2$ respectively, such that V$($$T_1$$)$ $\cap$ V$($$T_2$$)$ = $\phi$. Let u $\in$ V$($$T_1$$)$ and v $\in$ V$($$T_2$$)$. Let T be a tree obtained from $T_1$ and $T_2$ by introducing the edge uv. Then
\begin{align}
W(T) = W(T_1) + W(T_2) + n_1D_{T_2}(v) + n_2D_{T_1}(u) + n_1n_2.
\end{align}
\end{cor}
\indent Using Lemma~\ref{lem1}, we obtain an expression for $W$($T$), where $T$ is a binomial tree. It is known \cite{dob, dank} that for a tree $T$ with $\mid$$V$($T$)$\mid$ = $n$, $W$($T$) can be computed by an algorithm in time $O$($n$); thus for Fibonacci tree $T_{f_k}$ and binary Fibonacci tree $T^{b}_{f_k}$ with $\mid$$V$($T_{f_k}$)$\mid$ = $F_{k+2}$ and $\mid$$V$($T^{b}_{f_k}$)$\mid$ = $F_{k+2}$-1 respectively, such an algorithm can compute $W$($T_{f_k}$) or $W$($T^{b}_{f_k}$) in time $O$($F_{k+2}$) and requires as an input a description of $T_{f_k}$ or $T^{b}_{f_k}$, e.g., as adjacency list. Using Corollary~\ref{cor1}, we outline computational procedures for computing $W$($T_{f_k}$) and $W$($T^{b}_{f_k}$) in time $O$($\log{F_k}$), assuming that the input is only the order $k$ of the tree.
\section{Wiener index of binomial trees}
\begin{myth}
\it Let $T_{B_k}$ be a binomial tree with $2^k$ nodes. Then the Wiener index W$($$T_{B_k}$$)$ is given by \rm
\begin{center}
$W$($T_{B_k}$) = ($k$ $-$ 1)$2^{2k-1}$ + $2^{k-1}$.
\end{center}
\end{myth}
\em Proof. \rm Consider the binomial tree of Fig.~\ref{fig1}(b). For a vertex $r$ $\in$ $T^{''}_{B_{k-1}}$, let ${D}_{T^{'}_{B_{k}}}$($r$) be the sum of the distances of the form $d$($r$, $x$), for all $x$ $\in$ $T^{'}_{B_{k-1}}$. For convenience we denote ${D}_{T^{'}_{B_{k}}}$($r$) = $D_{T^{'}_B}$($k$). We first derive a closed form expression for $D_{T^{'}_B}$($k$) and $D_{T^{''}_B}$($k$) and then derive an expression for $W$($T_{B_k}$). From Fig.~\ref{fig1}(b), we have
\begin{align}
\label{four}
D_{T^{'}_B}(k) & = 2^{k-1} + D_{T^{'}_B}(k-1) + D_{T^{'}_B}(k-2) + \ldots + D_{T^{'}_B}(1). 
\end{align}
Also
\begin{align}
\label{five}
D_{T^{'}_B}(k-1) & = 2^{k-2} + D_{T^{'}_B}(k-2) + D_{T^{'}_B}(k-3) + \ldots + D_{T^{'}_B}(1).
\end{align}
>From \eqref{four} and \eqref{five}
\begin{align}
\label{six}
D_{T^{'}_B}(k) & = 2^{k-2} + 2D_{T^{'}_B}(k-1).
\end{align}
Solution to \eqref{six} yields
\begin{align}
\label{seven}
D_{T^{'}_B}(k) = (k + 1)2^{k-2}.
\end{align}
Similar to \eqref{four} we have 
\begin{align}
\label{eight}
D_{T^{''}_B}(k) = D_{T^{'}_B}(k) - 2^{k-1}. 
\end{align}
Using \eqref{seven} in \eqref{eight} we get
\begin{align}
\label{nine}
D_{T^{''}_B}(k) = (k + 1)2^{k-2} - 2^{k-1} = (k - 1)2^{k-2}.
\end{align}
Applying Lemma~\ref{lem1} to Fig.~\ref{fig1}(b) we get
\begin{align}
\label{ten}
W(T_{B_k}) = W(T^{'}_{B_{k-1}}) + W(T^{''}_{B_{k-1}}) + 2^{k-1}\big[D_{T^{'}_B}(k)\big] + 2^{k-1}\big[D_{T^{''}_B}(k)\big]. 
\end{align}
Using \eqref{seven} and \eqref{nine} in \eqref{ten} we get
\begin{align}
\label{eleven}
W(T_{B_k}) & = W(T^{'}_{B_{k-1}}) + W(T^{''}_{B_{k-1}}) + 2^{k-1}\big[(k + 1)2^{k-2} + (k - 1)2^{k-2}\big]\\
\label{twelve}
 & = W(T^{'}_{B_{k-1}}) + W(T^{''}_{B_{k-1}}) + 2^{k-1} k 2^{k-1}. 
\end{align}
Since $T^{'}_{B_{k-1}}$ and $T^{''}_{B_{k-1}}$ are isomorphic to $T_{B_{k-1}}$, from \eqref{twelve} we get
\begin{align}
\label{t13}
W(T_{B_k})  = 2 W(T_{B_{k-1}}) + k2^{2k-2}.
\end{align}
Solution to \eqref{t13} gives the expression for $W$($T_{B_k}$) as given in the Theorem 1. \qed
\section{Computing Wiener index of Fibonacci trees}
\subsection{Fibonacci trees}
Consider the Fibonacci tree $T_{f_k}$ of Fig.~\ref{fig2}. For a vertex $r$ $\in$ $T_{f_k}$, let $D_{T_{f_k}}$($r$) be the sum of distances of the form $d$($r$, $x$), for all $x$ $\in$ $T_{f_k}$. For convenience we denote $D_{T_{f_k}}$($r$) = $D_{T_f}$($k$). In computing $W$($T_{f_k}$), we first derive a closed form expression for $D_{T_f}$($k$). From Fig.~\ref{fig2}(b), we have
\begin{align}
\label{t14}
D_{T_f}(k) = D_{T_f}(k-1) + D_{T_f}(k-2) + F_k,
\end{align}
with initial conditions $D_{T_f}$(0) = 0, $D_{T_f}$(1) = 1, and $D_{T_f}$(2) = 2.\\
We introduce the generating function $G$($z$) as
\begin{align}
\label{t15}
G(z) = D_{T_f}(1)z + D_{T_f}(2)z^2 + D_{T_f}(3)z^3 + D_{T_f}(4)z^4 + \ldots \;.
\end{align}
>From \eqref{t14} and \eqref{t15} it follows that
\begin{align}
\label{t16}
(1 - z - z^2)\;G(z) & = F_1z + F_2z^2 + F_3z^3 + F_4z^4 + \ldots \;= \frac{z}{1 - z - z^2}.
\end{align}
>From \eqref{t16} we get
\begin{align}
\label{t17}
G(z) = \frac{z}{(1 - z - z^2)^2} = \frac{1}{5z} \Bigg[\frac{1}{(1-\phi z)^2} + \frac{1}{(1-\hat{\phi}z)^2} - \frac{z}{(1 - z - z^2)} \Bigg].
\end{align}
>From \eqref{t17} it follows (see for example \cite{knuth}) that
\begin{align}
\label{18}
[z^k]G(z) = D_{T_f}(k) = \sum_{j=1}^{k+1} F_j F_{k-j+1} = \frac{1}{5}\Bigg[kF_{k+2} + (k+2)F_k\Bigg]. 
\end{align}
Now applying Corollary~\ref{cor1} to Fig.~\ref{fig2}(b) we have the expression for $W$($T_{f_k}$) as given in Lemma 2.
\begin{mylm}
\it Let $T_{f_k}$ be the Fibonacci tree of order $k$. Then the Wiener index W$($$T_{f_k}$$)$ is given by the following expression:\rm
\begin{align}
\label{t19}
W(T_{f_k}) = W(T_{f_{k-1}}) + W(T_{f_{k-2}}) + F_{k+1}D_{T_f}(k-2) + F_kD_{T_f}(k-1) + F_{k+1}F_k.
\end{align}
\end{mylm}
We now compute $W$($T_{f_k}$) as shown in Fig.~\ref{fig4}.\\
\begin{figure}[tb]
\begin{description}
\item[Function] \sc ~WI-Fib-Tree\rm($k$: an order of $T_{f_k}$) \\\\
// Computes $W$($T_{f_k}$).
\end{description}
\begin{enumerate}
\item \hspace{2mm} $\forall$ 1 $\leq$ $i$ $\leq$ $k+1$, Compute $F_i$;
\item \hspace{2mm} $W_1$ := 1; $W_2$ := 4;
\item \hspace{2mm} \bf for \rm $i$ := 3 \bf to \rm $k$ \bf do begin\rm
\item \hspace{12mm} $D_1$ := 0.2 \big(($i-1$) $F_{i+1}$ + ($i+1$) $F_{i-1}$\big); \;\;// by (18)
\item \hspace{12mm} $D_2$ := 0.2 \big(($i-2$) $F_{i}$ + ($i$) $F_{i-2}$\big);  \;\;// by (18)
\item \hspace{12mm} $W$ := $W_2$ + $W_1$ + $F_{i+1}$$D_2$ + $F_i$$D_1$ + $F_{i+1}$$F_i$;  \;\;// by (19)
\item \hspace{12mm} $W_1$ := $W_2$; 
\item \hspace{12mm} $W_2$ := $W$;
\item \hspace{2mm} \bf end; \rm
\item \hspace{2mm} \bf return $W$; \rm
\caption{Computing $W$($T_{f_k}$).}
\label{fig4}
\end{enumerate}
\end{figure}
We thus have the following result.
\begin{myth}
\it For a tree $T_{f_k}$ $($k $>$ 2$)$, we can algorithmically compute W$($$T_{f_k}$$)$ in time O$($$\log{F_k}$$)$. The input to the algorithm requires only the order $k$ of the tree $T_{f_k}$. \rm 
\end{myth}
\em Proof. \rm It suffices to note that in Fig. 4, step 1 requires $O$($\log{F_k}$) additions and steps 4-6 require $O$($\log{F_k}$) multiplications and additions. \qed 
\subsection{Binary Fibonacci trees}
In computing $W$($T^{b}_{f_k}$), we first derive a closed form expression for $D_{T^{b}_f}$($k$). Note that $D_{T^{b}_f}$(0) = 0, $D_{T^{b}_f}$(1) = 0, and $D_{T^{b}_f}$(2) = 1. From Fig.~\ref{fig3}, we have, 
\begin{align}
\label{t20}
& D_{T^{b}_f}(k) = D_{T^{b}_f}(k-1) + D_{T^{b}_f}(k-2) + F_{k+2} - 2.
\end{align}
In a manner similar to that in Section 3.1, we obtain
\begin{align}
\label{t21}
D_{T^{b}_f}(k) = \sum_{j=2}^{k+1} (F_{j+2} - 2) F_{k-j+1}.
\end{align}
Simplifying the right hand side of \eqref{t21} we get
\begin{align}
\label{t22}
D_{T^{b}_f}(k) = \frac{1}{5}(k-3)F_{k+3} + \frac{2}{5}(k-2)F_{k+2} + 2.
\end{align}
In Fig.~\ref{fig3}(b), we identify $T^{b}_{f_{k-1}}$ together with the edge ($r$, $r^{'}$) $\in$ $T^{b}_{f_k}$ as $T^{b}_{f_{k-1}}$. Now applying Corollary~\ref{cor1} to Fig.~\ref{fig3}(b) we have the expression for $W$($T^{b}_{f_k}$) as given in Lemma 3.
\begin{mylm}
\it Let $T^{b}_{f_k}$ be the binary Fibonacci tree of order $k$. Then the Wiener index W$($$T^{b}_{f_k}$$)$ is given by the following expression:\rm
\begin{align*}
W(T^{b}_{f_k}) = W(T^{b}_{f_{k-1}}) & + W(T^{b}_{f_{k-2}}) + F_{k+1}D_{T^{b}_f}(k-2)\\
& + (F_k - 1)D_{T^{b}_f}(k-1) + F_{k+1}(F_k - 1).\tag{23}
\end{align*}
Using (23) we can compute $W$($T^{b}_{f_k}$) similar to the code fragment in Fig.~\ref{fig4}.
\end{mylm}

\end{document}